\documentclass[preprint,aps]{revtex4}
\usepackage[dvips]{epsfig}
\begin{document}

\title{Zero permeability and zero permittivity band gaps in 
1D metamaterial photonic crystals}
\author{Ricardo A. Depine \footnote{email: rdep@df.uba.ar}}
\altaffiliation{Member of CONICET 
(Consejo Nacional de Investigaciones Cient\'{\i}ficas y Tecnol\'ogicas).}
\affiliation{Grupo de Electromagnetismo Aplicado,
Departamento de F\'{\i}sica,
Facultad de Ciencias Exactas y Naturales, 
Universidad de Buenos Aires, 
Ciudad Universitaria, Pabell\'{o}n I, 
C1428EHA Buenos Aires, Argentina}
\author{Mar\'{\i}a L. Mart\'{\i}nez-Ricci \footnote{email: mricci@df.uba.ar}}
\affiliation{Grupo de Electromagnetismo Aplicado,
Departamento de F\'{\i}sica,
Facultad de Ciencias Exactas y Naturales, 
Universidad de Buenos Aires, 
Ciudad Universitaria, Pabell\'{o}n I, 
C1428EHA Buenos Aires, Argentina}
\author{Juan A. Monsoriu \footnote{email: jmonsori@fis.upv.es}}
\affiliation{Departamento de F\'{\i}sica Aplicada, 
Universidad Polit\'ecnica de Valencia, 
46022 Valencia, Spain}
\author{Enrique Silvestre \footnote{email: enrique.silvestre@uv.es}}
\affiliation{Departamento de \'Optica, Universidad de Valencia, 
46100 Burjassot, Spain}
\author{Pedro Andr\'es \footnote{email: pedro.andres@uv.es}}
\affiliation{Departamento de \'Optica, Universidad de Valencia, 
46100 Burjassot, Spain}
\date{\today}
\pacs{42.70.Qs, 78.20.Ci, 41.20.Jb, 78.67.Pt}
\keywords{photonic crystals, negative refraction, metamaterials, negative index}
\begin{abstract}
We consider layered heterostructures combining ordinary positive index materials and dispersive metamaterials. 
We show that these structures can exhibit a new type of photonic gap near frequencies where either the magnetic permeability $\mu$ or the electric permittivity $\epsilon$ of the metamaterial change signs. 
Although the interface of a medium with zero refractive index (a condition attained either when 
$\mu=0$ or when $\epsilon=0$) is known to give full reflectivity for all incident polarizations, here we show that a gap corresponding to $\mu=0$ occurs only for TE polarized waves, whereas a  gap corresponding to $\epsilon=0$ occurs only for TM polarized waves. These band gaps are scale-length invariant and very robust against disorder, although they may disappear for the particular case of propagation along the stratification direction. 
\end{abstract}

\maketitle

Photonic band gap (PBG) materials allow electromagnetic field propagation in certain frequency bands, but not in others, their essential feature being the periodic arrangement of high-contrast electromagnetic properties. The simplest version of a PBG material is the 1D planar layered stack, which has been commonly used in optics in the form of filters or Bragg reflectors \cite{yeh1}. 

Almost every available natural material has been used to construct PBG structures. Possibilities have been widened with the recent advent of metamaterials (MMs), artificially constructed composites exhibiting electromagnetic properties that are difficult or impossible to achieve with conventional, naturally occurring materials \cite{SmithScience04}. 
Key representatives of this new class of materials are MMs with negative index of refraction, a property that arises in media with a negative electric permittivity together with a negative magnetic permeability in the same frequency range \cite{Veselago}. 
Multilayered PBG structures containing MMs have attracted much attention in recent years 
\cite{gl1,liprl90,SNG1,completeshadrivov}. Two new kinds of PBG, fundamentally different from the usual Bragg gaps originating from interference in the periodic structure, have been identified: the ''zero averaged refractive index'' gap \cite{liprl90} and the ''zero effective phase'' gap \cite{SNG1}. The first one appears in multilayers combining ordinary materials (positive refractive index) and MMs with negative refractive index, whereas the second one appears in multilayers containing two different single-negative (permittivity- or permeability-negative) MMs. In contrast to Bragg PBGs, both the ''zero averaged refractive index'' and the ''zero effective phase'' gaps remain invariant to scale-length changes and are robust against disorder. PBGs with these characteristics 
can be  useful as photonic barriers in quantum well applications, such as multiple channeled filtering \cite{SNG1}. 

Since Bragg PBGs result from interference between waves reflected at different layers, material dispersion does not play a crucial role in their appearence. In other words, Bragg PBGs in 1D multilayers always exist, even when the dispersion of the constituent materials is neglected. 
On the other hand, the existence of a ''zero averaged refractive index'' PBG highly relies on the dispersive characteristics of the material layers, since for nondispersive materials this PBG would cover  all frequencies, except for singular points \cite{liprl90}. In this Letter we show that MM multilayers can also exhibit a new kind of PBG, where material dispersion --and not interference-- plays the key role. These new PBGs correspond to frequency regions where the constitutive parameters of the MM change their signs. Since they are not based on the Bragg interference mechanism, they are also invariant to scaling and even more insensitive to disorder 
than the ''zero averaged refractive index'' and the ''zero effective phase'' gaps.

If losses are neglected, the refractive index of a MM $n(\omega)=\sqrt{\mu(\omega) \epsilon(\omega)}$ is zero at the values of the frequency $\omega$ where the magnetic permeability $\mu(\omega)$ or the electric permittivity $\epsilon(\omega)$ change their sign. Therefore, at these frequency values one could expect to find gaps for both TE and TM polarization, since it is well known that the boundary of a zero refractive index material exhibits full reflectivity \cite{garciaAPL,ultralow}. In contrast to this expectation, we show that 
at frequencies where $\mu=0$ a gap occurs for TE polarized waves, but not necesarily for TM polarized waves, whereas at frequencies where $\epsilon=0$ a gap occurs for TM polarized waves, but not necessarily for TE polarized waves. Moreover, we show that these gaps can be completely absent for the particular case of propagation along the stratification direction, which could explain why its presence went unnoticed in previous studies (see, for example, Fig. 2 in \cite{liprl90}). 

We consider the 1D periodic structure created by layers of two different materials: a conventional dielectric with permeability $\mu_1$, permittivity $\epsilon_1$ and thicknesses 
$d_1$, and a MM with permeability $\mu_2(\omega)$, permittivity $\epsilon_2(\omega)$ and thicknesses $d_2$. The stratification direction is the $y$ axis and we consider wave propagation in the $x-y$ plane. Let the function $f(x, y)$ represent the $z$--directed component of the electric field for the TE--polarization case (electric field parallel to the layers), and the $z$--directed component of the magnetic field for the TM--polarization case (magnetic field parallel to the layers). The propagating waves in the periodic structure have the form of Bloch modes, for which the fields satisfy the condition $f(x, y+d)=f(0, y)\exp{i(k_x x + K d)}$, where $d=d_1+d_2$ is the period of the structure, $k_x$ is the wave vector component along the layers and $K$ is the Bloch wave number. For two-layered periodic structures, the dispersion relation 
$K(\omega,k_x)$ can be found explicitly from \cite{yeh1}
\begin{eqnarray}
\cos(Kd)=\xi \equiv \cos(k_{1y} d_1)\,\cos(k_{2y} d_2) - \,\,\,\,\,\,\,\,\,\,\,\,\,\nonumber \\ 
\frac{1}{2}\left[
\frac{\sigma_2 k_{1y}}{\sigma_1 k_{2y}} + 
\frac{\sigma_1 k_{2y}}{\sigma_2 k_{1y}}\right]
\sin(k_{1y} d_1)\,\sin(k_{2y} d_2) \,\,, \label{dispersion}
\end{eqnarray} 
where the index $j = 1,2$ indicates the layer, $\sigma_j=\mu_j$ for TE polarization or 
$\sigma_j=\epsilon_j$ for TM polarization, $k_{jy}^2=k_j^2-k_x^2$, and $k_j=\omega n_j/c$ are wave numbers in each media with refractive indexes $n_j$. The quantity $\xi$, half the trace of the matrix characterizing the unit cell translation operator \cite{yeh1}, determines the band structure. It takes real values for lossless media and real $k_x$. Regimes where $|\xi| < 1 $ correspond to real $K$ and thus to propagating Bloch waves. In regimes where $|\xi| > 1$, $K$ has an imaginary part, therefore the Bloch wave is evanescent, and this regime corresponds to forbidden bands (or gaps) of the periodic medium. The band edges are those regimes where $|\xi| = 1$. 

Let us consider frequency regions where the refractive index $n_2(\omega)$ goes through zero, that is, where a constitutive parameter of the MM --either $\mu_2(\omega)$ or 
$\epsilon_2(\omega)$-- changes its sign. Note that the term between square brackets in the right hand side of eq. (\ref{dispersion}) becomes singular under these conditions. 
For propagation along the stratification direction ($k_x=0$), $k_{2y} \rightarrow 0$ and the singularity is compensated. To be specific, and taking into account that $\mu_2$ and $\epsilon_2$ do not generally change their sign at the same frequency value, we assume that $\epsilon_2 \rightarrow 0$ but $\mu_2 \neq 0$. In this limit, $k_{2y} \rightarrow 0$ and the quantity $\xi$ for both TE and TM polarizations adopts the form
\begin{eqnarray}
\xi \approx \cos(k_{1} d_1) -  \frac{1}{2} \frac{\omega}{c} 
\sqrt{\frac{\epsilon_1}{\mu_1}}\, \mu_2 d_2 \;\sin(k_{1} d_1) \,\,, \label{displim1}
\end{eqnarray}
which can take any real value, either outside or inside the interval $\left[-1,1\right]$. A similar  
conclusion can be obtained when $\mu_2 \rightarrow 0$ but $\epsilon_2 \neq 0$. 
Thus, gaps near regions where a constitutive parameter changes its sign do not occur  automatically at normal incidence. 
For example, the multilayer considered to obtain 
Fig. 2 in Ref. \cite{liprl90}, exhibits Bragg gaps and a ''zero averaged refractive index'' gap, but there is no gap in the regions where $\mu_2$ and $\epsilon_2$ change signs. This can be observed in the band diagrams shown in Figs. 1b and 1f, obtained for $k_x=0$ and the following geometric and constitutive parameters \cite{liprl90}: 
$d_1=12$ mm, $d_2=6$ mm, $\epsilon_1=\mu_1=1$, 
\renewcommand{\arraystretch}{2}
\begin{equation}
\begin{array}{l}
\epsilon_{2}(f)=1 + \frac{\displaystyle 5^2}{\displaystyle 0.9^2-f^2}+
\frac{\displaystyle 10^2}{\displaystyle 11.5^2-f^2}\,,\\
\mu_{2}(f)=1 + \frac{\displaystyle 3^2}{\displaystyle 0.902^2-f^2}\,,
\end{array}
\label{eq13}
\end{equation}
\renewcommand{\arraystretch}{1}
where $f$ is the frequency measured in GHz. 

For oblique incidences, on the other hand, $k_x \neq 0$, $k_{2y} \rightarrow \pm i k_x \neq 0$ and the second term in the right hand side of eq. 1 remains singular when $\mu_2=0$ (TE polarization) or when $\epsilon_2=0$ (TM polarization), thus originating new gaps, as shown in Fig. 1 for $\theta=30^\circ$ and $60^\circ$ ($k_x=k_1 \sin \theta$). The non singular behavior 
of $\xi$ at normal incidence and its singular behavior for oblique incidences is illustrated in Fig. \ref{trazas} for both polarizations. 
We conclude that these new gaps always appear for non zero propagation angles $\theta$ in layer 1: the $\mu_2=0$ gap appears for TE, but not for TM, polarization; whereas the $\epsilon_2=0$ gap appears for TM, but not for TE, polarization. 

The projected band structure corresponding to this example is shown in Fig. \ref{projected}. White regions indicate forbidden bands where there are no electromagnetic modes, regardless of $K$. The width of the ''zero averaged refractive index'' gap does not change appreciably for TE polarization,  but it gets narrower when $\theta \rightarrow 90^\circ$ for TM polarization. 
On the other hand, the 
$\mu_2=0$ and the $\epsilon_2=0$ gaps increase with the angle $\theta$. 

Like the ''zero averaged refractive index'' gap \cite{liprl90} and the ''zero effective phase'' gap \cite{SNG1}, the gap arising from $\mu_2=0$ or $\epsilon_2=0$ differs fundamentally from the usual Bragg gap. First, its central frequency is independent of the lattice constant, while all Bragg gap frequencies must scale with the lattice constant. This fact is shown in Fig. \ref{escalados}, where we compare the transmittance through 16 unit cells corresponding to the periodic multilayer considered in Fig. \ref{projected} for $\theta=45^\circ$, with the transmittance obtained for similar structures with lattice constants scaled by factors of $2/3$ and $4/3$. While the Bragg gap shifts upward or downward in frequency, the ''zero averaged refractive index'' gap --near 2.3 GHz (TE) and 3.55 GHz (TM)-- and the new gaps --near 3.55 GHz (TE) and 3.9 GHz (TM)-- remain unchanged. Second, the Bragg gap, relying on interference mechanisms, is destroyed by strong disorder, while the $\mu_2=0$ and the $\epsilon_2=0$ gaps, relying on constitutive properties of the MM, should be expected to be robust against disorder. In Fig. \ref{distorsiones} we compare the transmittance through 16 unit cells corresponding to the periodic multilayer considered in Fig. \ref{projected} for $\theta=45^\circ$ and to similar structures with thickness  fluctuation of $\pm 3$ mm and $\pm 6$ mm, each ensemble averaged over 24 realizations. As expected, the Bragg gap is destroyed by disorder, but the ''zero permeability'' and the ''zero permittivity'' gaps survive. 

To summarize, we have shown that periodic multilayers containing MMs can exhibit a new type of photonic gap near frequencies where either the magnetic permeability $\mu$ or the electric permittivity $\epsilon$ of the MM change signs. In contrast to Bragg PBGs, originating from interference mechanisms, it is the material dispersion of the MM what mainly determines the appearence of these new gaps. Therefore, like the ''zero averaged refractive index'' and the ''zero effective phase'' gaps, the new gaps also remain invariant to scaling and insensitive to disorder. Although electromagnetic waves would be completely reflected at the boundary of a medium with zero refractive index, our results show that multilayers with MM constituents do not automatically exhibit PBGs at frequencies where the refractive index is zero: for propagation along the stratification direction, zero refractive index PBGs can be completely absent, whereas for oblique propagation they may emerge for different polarizations, depending on which of the constitutive parameters $\mu(\omega)$ or $\epsilon(\omega)$ makes the refractive index zero. \\


This work was funded by the Plan Nacional I+D+i (grant TEC2005-07336-C02-02/MIC), Ministerio de Educaci\'on y Ciencia, Spain, and FEDER, and the Generalitat Valenciana, Spain (grant Grupos03/227). 
RAD acknowledges financial assistance provided by the Universidad de Valencia (Programa de Estancias Temporales para Investigadores Invitados). 
MLMR acknowledges partial support from Consejo Nacional de Investigaciones Cient\'{\i}ficas y T\'ecnicas (CONICET), Universidad de Buenos Aires (UBA) and Agencia Nacional de Promoci\'on Cient\'{\i}fica y Tecnol\'ogica (ANPCYT-BID 802/OC-AR03-14099).

\newpage
\thispagestyle{empty}
\begin{figure}[tp]
\begin{center}
\includegraphics[width=12cm]{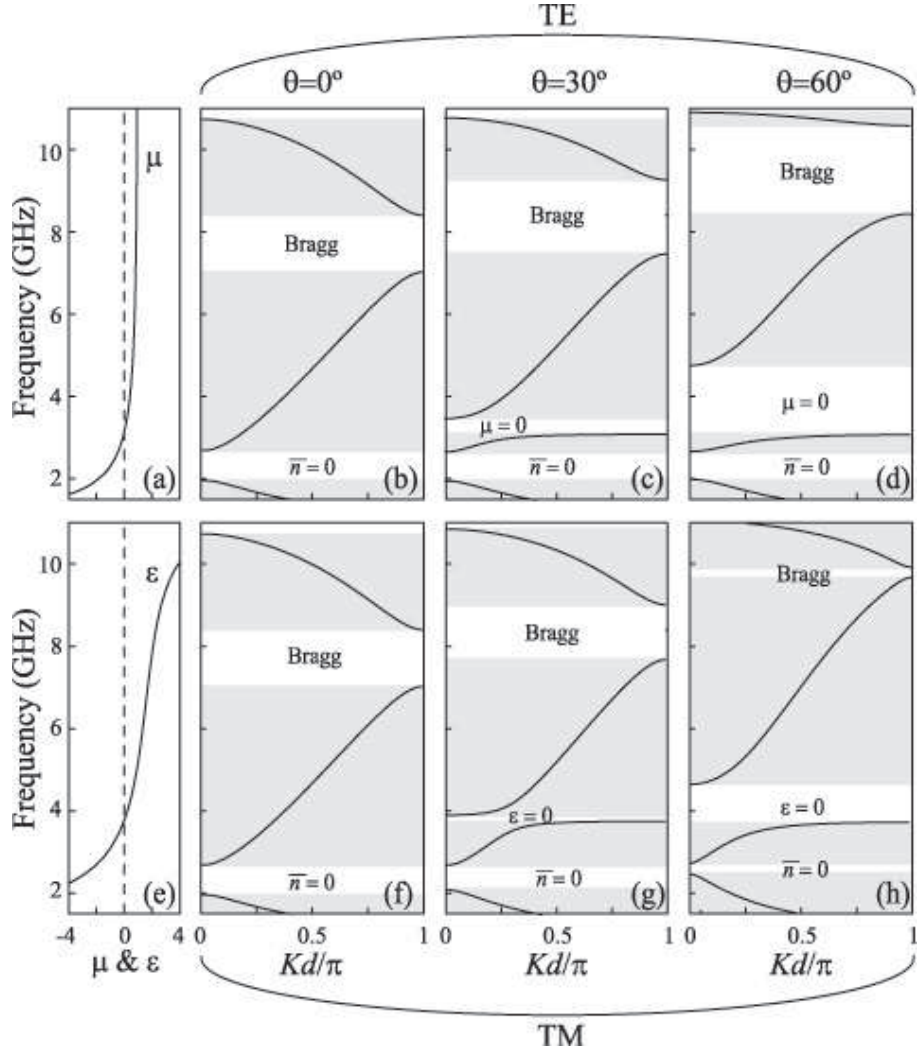}
\end{center} 
\caption{Band structure for TE and TM polarizations and different angles of incidence corresponding to a periodic stack with air layers ($\mu_1=\epsilon_1=1$, $d_1=12$ mm) and MM layers ($\mu_2$ and $\epsilon_2$ given by eq. \ref{eq13}, $d_2=6$ mm). The left column shows the frequency behavior of the constitutive parameters corresponding to layer 2. 
\label{bandgaps}}
\end{figure}

\newpage
\thispagestyle{empty}
\begin{figure}[tp]
\begin{center}
\includegraphics[width=12cm]{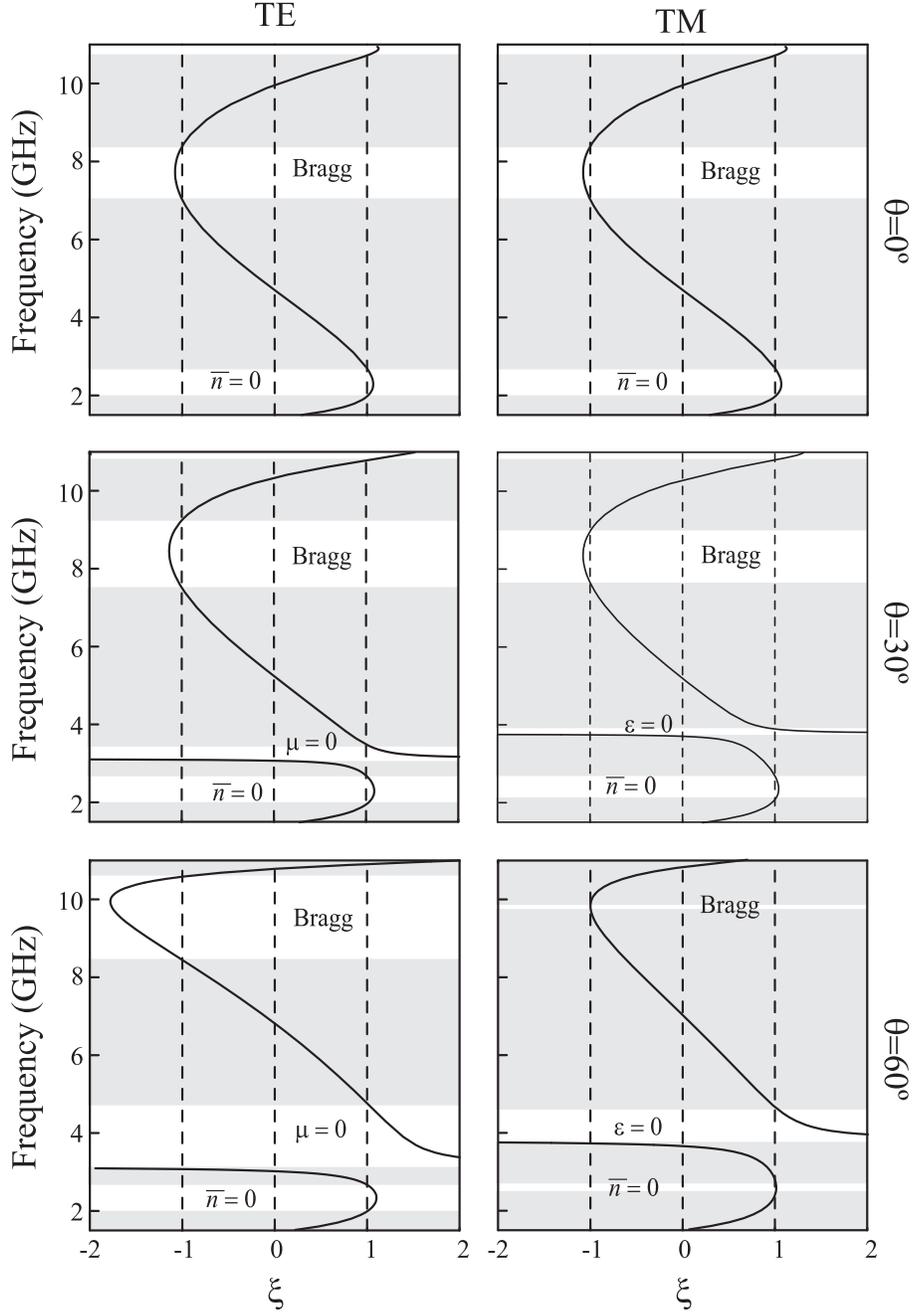}
\end{center} 
\caption{Half of the trace of the unit cell translation matrix, $\xi$, as given by eq. \ref{dispersion}, for the structure considered in 
Fig. \ref{bandgaps}, $\theta=0^\circ$, 
$30^\circ$, and $60^\circ$. The left column corresponds to TE polarization, whereas the right column corresponds to TM polarization.}\label{trazas}
\end{figure}

\newpage
\thispagestyle{empty}
\begin{figure}[tp]
\begin{center}
\includegraphics[width=12cm]{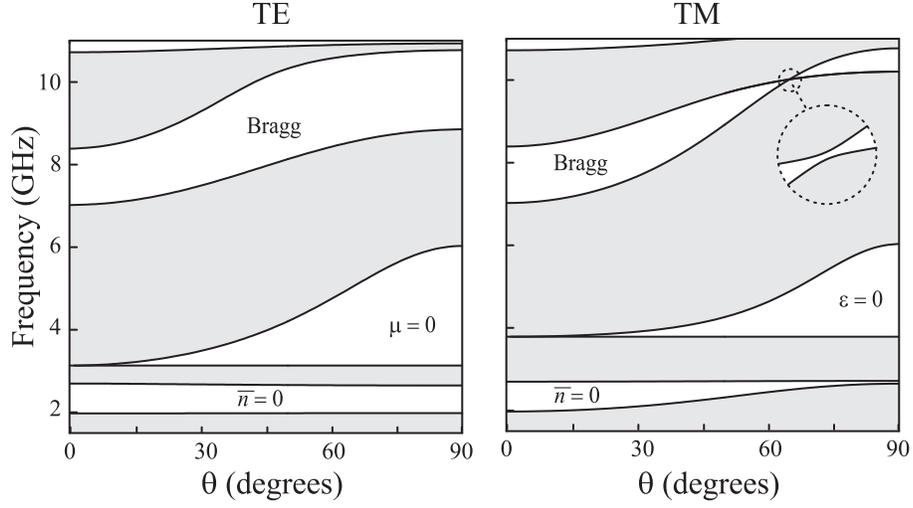}
\end{center} 
\caption{Projected band structure for the MM multilayer considered in 
Fig. \ref{bandgaps}.}\label{projected}
\end{figure}

\newpage
\thispagestyle{empty}
\begin{figure}[tp]
\begin{center}
\includegraphics[width=14cm]{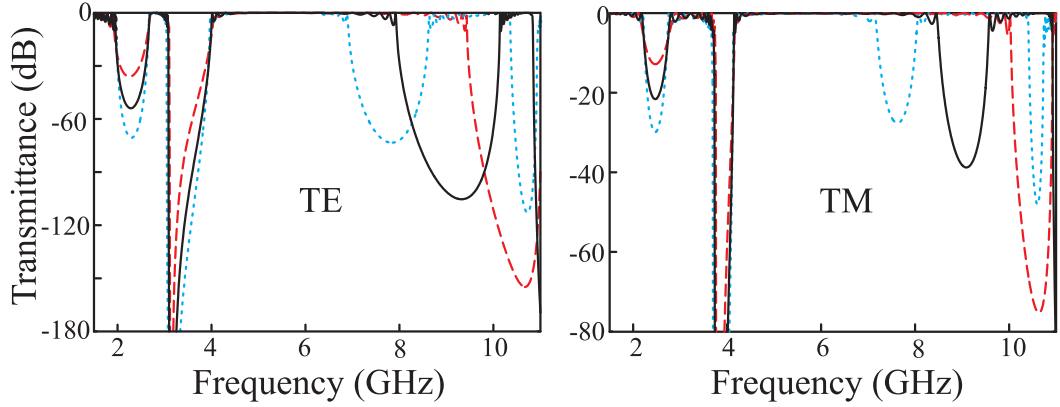}
\end{center} 
\caption{(color online) TE and TM transmittance through 16 unit cells corresponding to the band structure in 
Fig. \ref{projected} for $\theta=45^\circ$ (solid line), the same structure but the lattice constant scaled by $2/3$  (dashed line), and the same structure but the lattice constant scaled by $4/3$ (dotted line). }\label{escalados}
\end{figure}

\newpage
\thispagestyle{empty}
\begin{figure}[tp]
\begin{center}
\includegraphics[width=14cm]{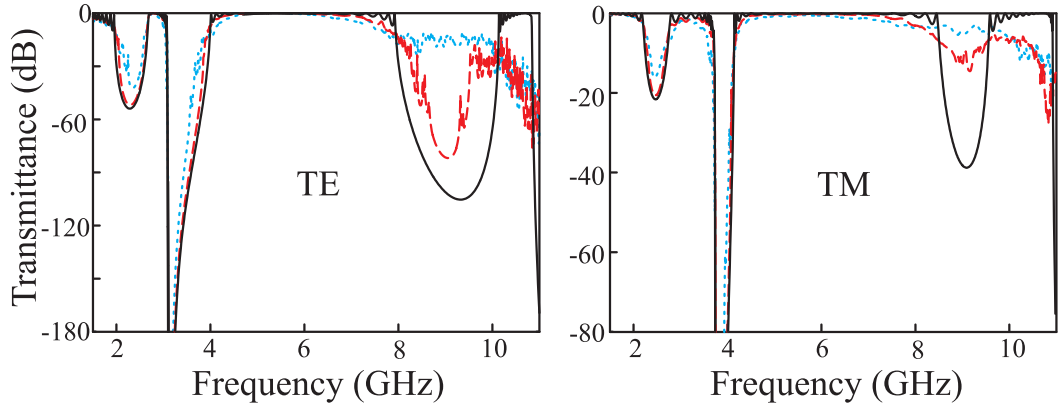}
\end{center} 
\caption{(color online) TE and TM transmittance through 16 unit cells corresponding to the perfectly periodic multilayer considered in Fig. \ref{projected} for $\theta=45^\circ$ (solid line), to a similar structure with thickness fluctuation of $\pm 3$ mm (dashed line), and thickness fluctuation of $\pm 6$ mm (dotted line), averaged over 24 samples. }\label{distorsiones}
\end{figure}

\end{document}